# Improving TMS–EEG Signal Quality for Closed-Loop Neurostimulation via Source-Domain Denoising

**Zhen Tang**[2], **Ameer Hamoodi**[1], **Stevie Foglia**[1], **Aimee Nelson**[1], **and Zhen Gao**[2]
[1]Department of Kinesiology, Faculty of Science, McMaster University, Canada
[2]Faculty of Engineering, McMaster University
* E-mails: tangz41@mcmaster.ca; nelsonaj@mcmaster.ca; gaozhen@mcmaster.ca

**Abstract:**

This research addresses a validated TMS–EEG cleaning pipeline and a corresponding benchmark dataset. It evaluates two widely used artifact removal pipelines. A reference dataset of carefully preprocessed EEG signals was established to support future algorithm development and enable systematic comparison of automated artifact removal strategies, despite the absence of a true physiological ground truth. The study evaluates the effectiveness of two widely used source-based artifact removal approaches and examines their impact on signal quality improvement and preservation of TMS-evoked potentials. The results support the robustness of the proposed preprocessing workflow and demonstrate its potential for improving data reliability in both research and clinical applications. A key goal is integrating TMS-EEG and embedding it within a larger BCI framework. Ultimately, these efforts aim to enhance understanding of cortical dynamics and expand the clinical and research applications of TMS-EEG.

## 1. Introduction

A major advantage of brain-computer interface (BCI) technology is its ability to monitor brain states in real time. When integrated with traditional brain stimulation techniques, it enables personalized, closed-loop interventions rather than fixed open-loop protocols. However, deriving the state of brain circuit from a few tens of seconds of an EEG signal is not only challenging due to the presence of noise in the EEG (such as ocular and muscle artifacts) but also due to a "curse of dimensionality" (Altman and Krzywinski, 2018) that makes the identification of a reliable mapping from a segment of TMS-EEG data (a matrix consisting of time by channel) to the activity of the circuit of interest at that time difficult.

Brain Stimulation has been expected to have the ability to reduce chronic pain by a significant amount, although many patients remain non-responders when using traditional open-loop systems. Incorporating a BCI enables the creation of a closed-loop system that continuously monitors real-time neural states and adjusts stimulation parameters accordingly, potentially enhancing therapeutic efficacy.

Neuronal electroencephalography (EEG) signals contain the necessary for determining brain activity. The physiological signals generally have amplitudes of around 5–10 uV, the system used to detect them is highly sensitive. (Ferree et al., 2001; Vrba and Robinson, 2001) Thus, it is very vulnerable to all types of artifacts from a wide range of sources. Among these, the most problematic artifacts include TMS-induced muscle activity, stimulator recharge transients, and somatosensory-evoked potentials. Several methods have been employed to isolate clean neuronal EEG responses. This work focuses on two source-based artifact-removal methods, covering their performance and theoretical background:

1. signal-space projection-based artifact suppression (Mutanen et al., 2022)
2. the source-estimate utilizing noise-discarding (SOUND) algorithm ((Mutanen et al., 2018)

The dataset was first preprocessed with a sample-based approach to maximize the performance of both algorithms and obtain cleaner data. Subsequently, it was subjected to Infomax independent component analysis (ICA), components corresponding to ocular, muscular, and cardiac activity were identified with ICLabel (Pion-Tonachini et al., 2019) and verified through visual inspection.

This study applied established TMS-EEG preprocessing techniques, including SSP-SIR, SOUND, and ICA, to an independent EEG dataset to evaluate artifact suppression effectiveness and preservation of neural signals. While these methods have been previously proposed and validated in prior literature, their performance may vary depending on recording conditions, stimulation parameters, and subject-specific characteristics. Therefore, systematic validation on independent datasets remains important. The contribution of this work is not the re-identification of known artifact types, but the implementation and validation of a practical preprocessing pipeline for real EEG recordings, with emphasis on artifact removal effectiveness, signal quality improvement, and reliable interpretation of TMS-evoked potentials.

In the first half of this research, we aimed to explore the sources and patterns of potential artifacts and develop strategies to eliminate or suppress them. By targeting primary motor cortex(M1) on the scalp, through EEG, we measured the pre-stimulus PS between the TMS target (M1) and the regions of secondary responses. The target area activated by the TMS represents the presynaptic neural group, and the TEP amplitude induced as a secondary response in distant areas provides the postsynaptic input gain. All artifacts are assumed to be independent.

In the second phase of this study, we aim to establish a reliable reference dataset of carefully preprocessed EEG signals and systematically explore different artifact removal methodologies. Since the Source Estimate Utilizing Noise Discarding (SOUND) algorithm has been proven effective (Mutanen et al., 2018). It was selected for further implementation in Python based on the open-source MATLAB toolbox TESA. This transition improves accessibility, reproducibility, and supports future integration into real-time closed-loop BCI systems.

## 2. Literature Review

The signal-space projection (SSP), a spatial filtering method initially proposed by Mäki and Ilmoniemi (2011), was designed to remove muscle artifacts caused by TMS stimulation. However, applying spatial filtering often alters the original spatial distribution of neuronal EEG signals, making it challenging to visually interpret the cleaned EEG data. To address this issue, Mutanen et al. (2016) introduced Source-Informed Reconstruction (SIR), a method aimed at restoring the original neuronal spatial patterns after artifact removal.

Building directly upon the principles of SIR, Mutanen et al. (2018) developed the Source-estimate Utilizing Noise-Discarding (SOUND) algorithm, which automatically identifies and eliminates EEG noise.

By cross-validating EEG signals, we can identify signal components that are more likely to originate from extracranial sources, i.e., noise. Incorporating an EEG forward model enhances the accuracy of this cross-validation based on the correlation patterns among EEG channels. This method is known as the Source-estimate-Utilizing Noise-Discarding algorithm, or SOUND (Mutanen et al., 2018).

EEG has poor spatial specificity due to the conductive properties of the scalp, neuronal EEG signals are dominated by topographical patterns with low spatial specificity, which lead the EEG signals to be highly correlated across the sensors, and nearby EEG channels often pick up similar voltage values.

## 3. Methods

Following diagram (Figure 1) illustrates the workflow of the TMS–EEG artifact removal pipeline. EEG data were collected using a 32-channel EasyCap system with TMS applied over the left primary motor cortex (M1). Preprocessing included filtering, downsampling, and artifact flagging to prepare the data for Infomax ICA with ICLabel classification, which removed ocular, muscle, and cardiac components. Signal-Space Projection (SSP) was applied to further suppress high-amplitude artifacts, followed by the Source-Estimate Utilizing Noise-Discarding (SOUND) algorithm for source-domain cleaning while preserving genuine neural activity. The final output was a benchmark ground truth dataset suitable for validating future TMS–EEG cleaning methods.

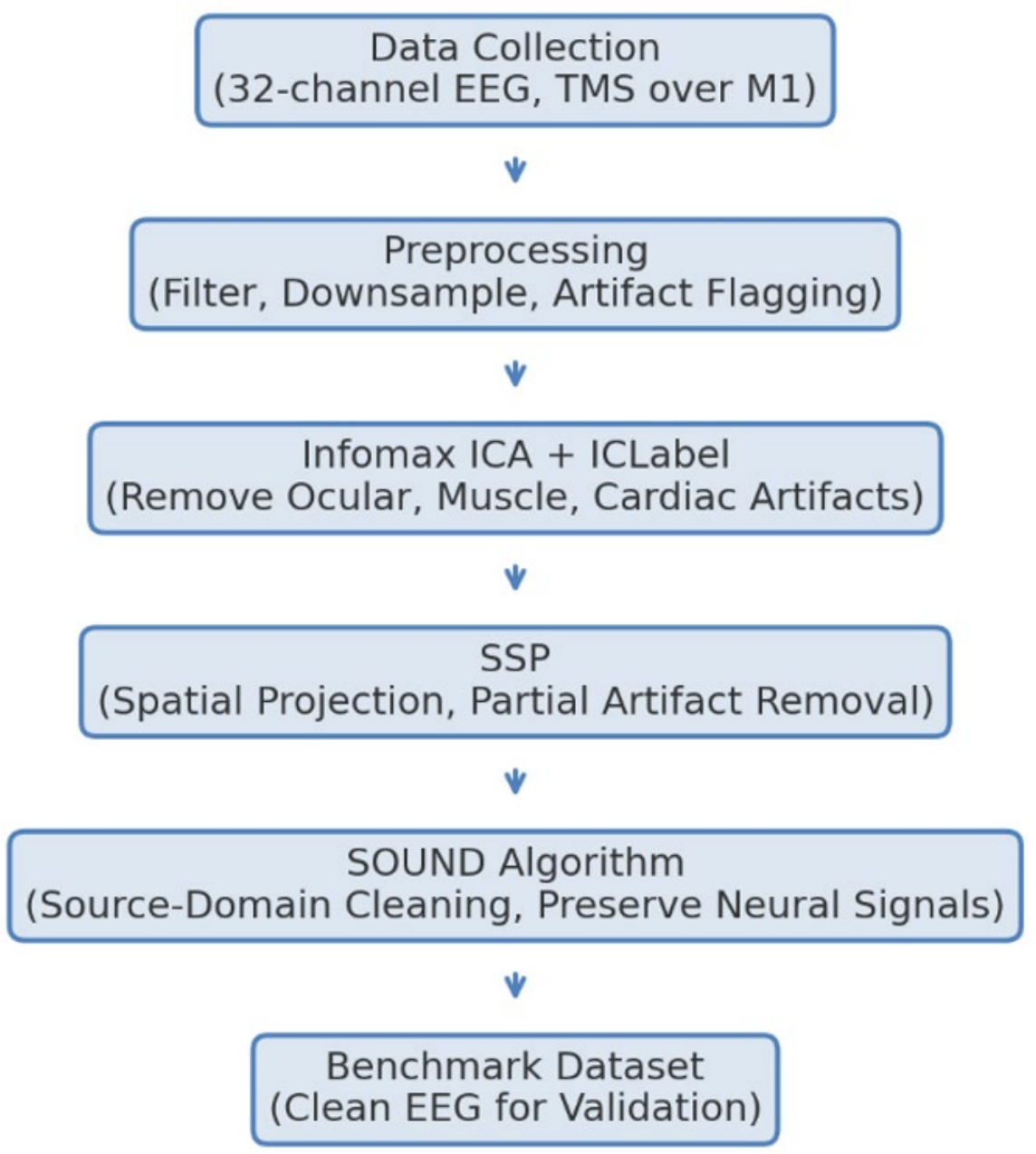

Figure 1 pipeline diagram

### 3.1 Data Collection and Processing

EEG data were collected using a 32-channel EasyCap system, with the TMS coil positioned over the left primary motor cortex (M1, C3) at a 45° angle to the midline, inducing a posterior–anterior current. The stimulation intensity was set at 100% of the resting motor threshold. A total of 50 trials were conducted, with single pulses delivered every 3 seconds, and markers were recorded for synchronization.

The preprocessing and event related desynchronization:

- Remove channels ±2 SD from mean SD.
- FIR filtering: high-pass 1 Hz, low-pass 40 Hz.
- Down-sample to 250 Hz.
- Reject points > ±500 μV.
- Segment into 1 s pseudo-epochs for ICA.
- Re-reference to common average (no spherical interpolation due to technical constraints).
- Analysis limited to 30 electrodes.

### 3.2 Infomax ICA

Preprocessed Data were segmented into 3s epochs and applied with infomax ICA for independent component analysis with 15 components. The results are as shown in Figure 2.

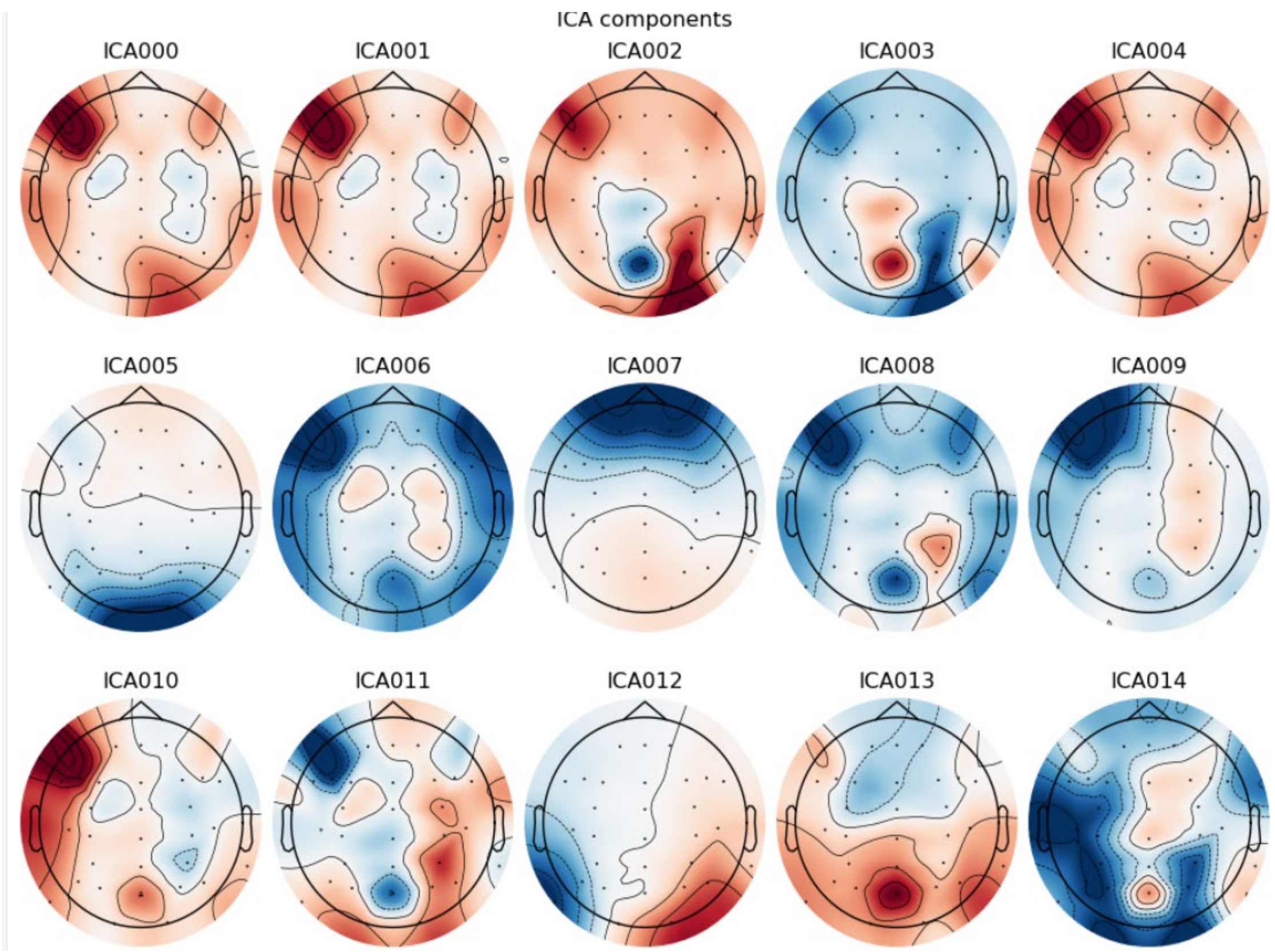

Figure 2 ICA topography

ICA effectively suppresses artifacts by projecting sensor data onto a sub-space that is orthogonal to the artefact pattern. However, this operation inevitably removes some neuronal activity whose spatial patterns overlap with the rejected sub-space. As a result, although the data becomes cleaner, their scalp topographies are distorted. To minimize such distortion, only components that are almost 100% artifactual should be removed at current stage.

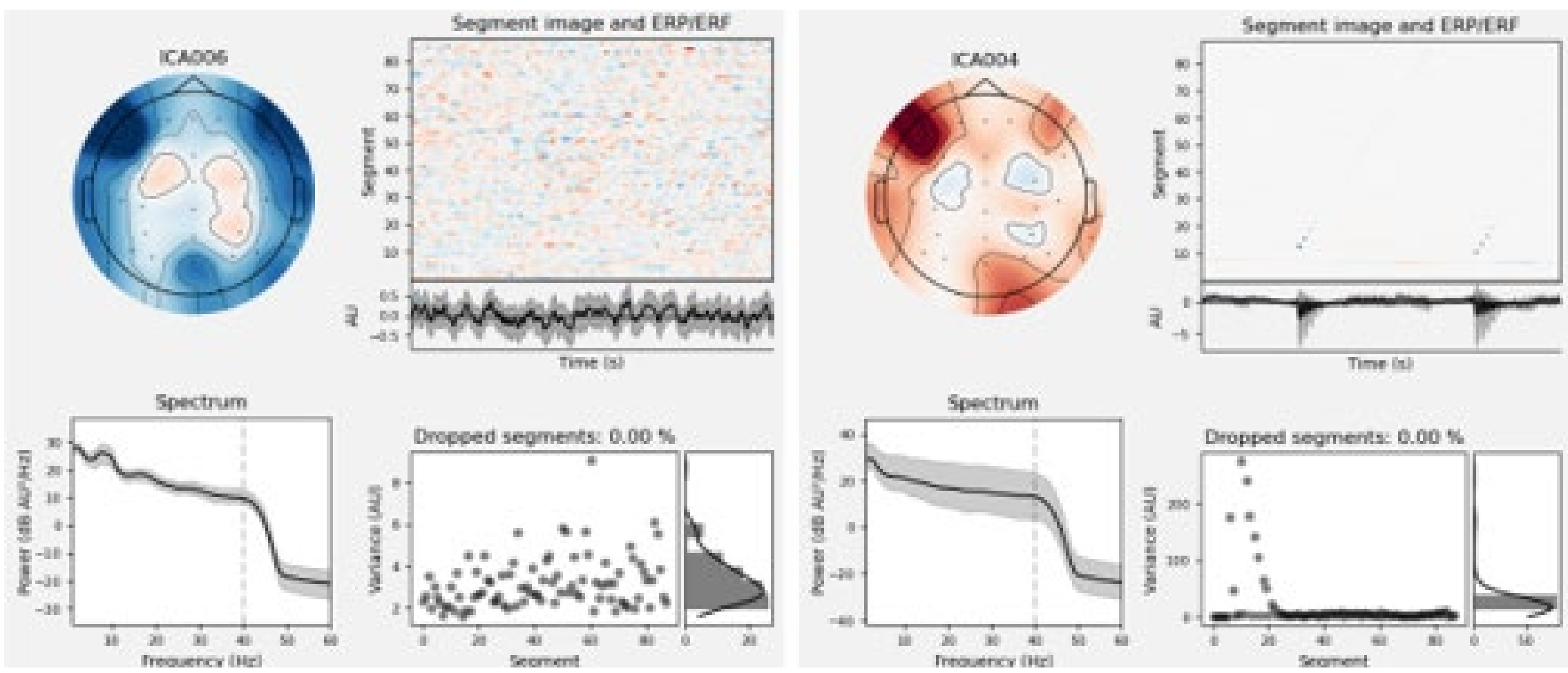

Figure 3 Brain source vs artifacts

### 3.3 Contaminated components exclusion using ICLabel

ICLabel sorts ICA components into seven categories: Brain components show a cortical dipolar scalp map, a 1/f spectrum with alpha-range peaks, and clean ERP structure; Eye components cluster over the eyes, dominate < 5 Hz, and contain blink or saccade transients; Muscle components are peripheral, highly focal, and exhibit broadband high-frequency bursts; Heart (ECG) components form a near-linear gradient across the scalp and display regular ~1 Hz QRS complexes; Line Noise components are identified mainly by a sharp 50 or 60 Hz peak in the power spectrum; Channel Noise components load almost entirely on a single electrode with large recurrent artefacts; and Other components lack these defining patterns, often appearing non-dipolar with weak or mixed spectral features (Altman & Krzywinski, 2018).

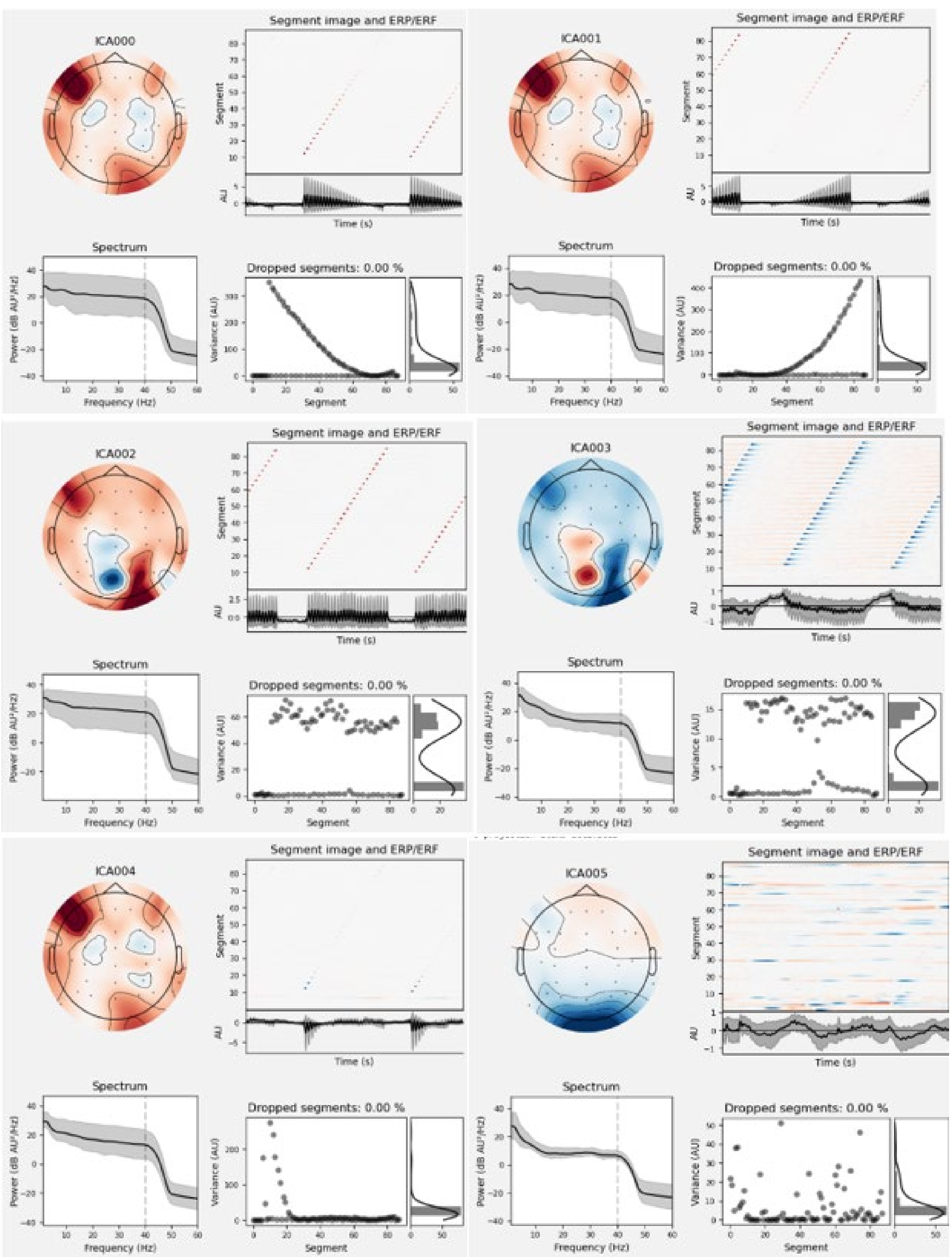

Figure 4 Artifact components, ICA0-4 as pure artifacts data since there is nothing in the ERP ICA 005 representing eye blinks

These six independent components clearly reflect artifactual activity. Their scalp topographies show abrupt, ring-like polarity transitions instead of the smooth dipolar distributions expected from cortical sources. The ERP images offer further confirmation, displaying large time-locked transients that bear no relation to experimental events. Similarly, their power spectra are dominated by broadband high-frequency activity and lack the characteristic around 10 Hz alpha peak of neuronal signals. Collectively, these signatures—spatially atypical maps, non-event-related temporal patterns, and non-physiological spectral profiles can identify the components as TMS-related muscle or hardware artifacts, leading to their exclusion from further analysis.

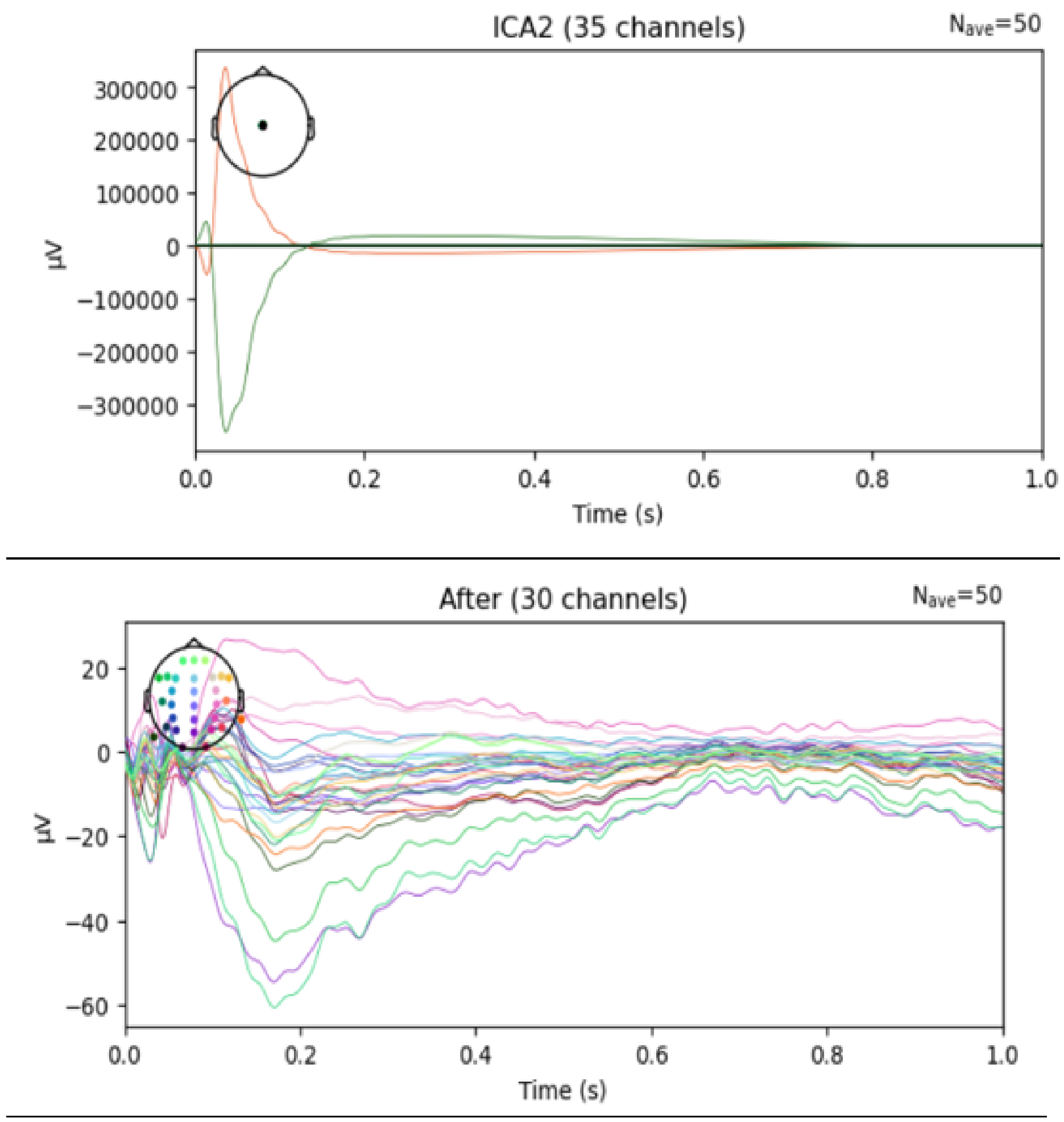

Figure 5 Informax ICA efficacy

Overlaying the 1s post-pulse traces shows that Infomax ICA removed most large-amplitude artifacts. The few residual artifacts appear only in isolated channels and are not time-locked across the array. The remaining waveforms share similar trends and amplitude, indicating that ICA suppressed noise without distorting the neural signals. The dataset is therefore suitable for the next cleaning steps.

### 3.4 Signal Space Projection Without Source-Informed Reconstruction

Signal Space Projection (SSP) is a spatial filtering method that can suppress high-amplitude artifacts by projecting EEG data away from artifact-related spatial components. In TMS-EEG preprocessing, SSP is commonly discussed together with Source-Informed Reconstruction (SIR), forming the SSP-SIR approach for suppressing TMS-evoked muscle artifacts (Mutanen et al., 2022). However, because artifact and neural signal may overlap, this process can inevitably distort the underlying neural signals. (Mutanen et al., 2022) To address this limitation, Source-Informed Reconstruction (SIR) should be applied after SSP to recover and preserve the original brain activity by reconstructing the neural sources that were suppressed during artifact removal. However, SIR requires a forward model built from structural MRI, which is unavailable for our lab trial. For this reason, only SSP was applied in the current analysis.

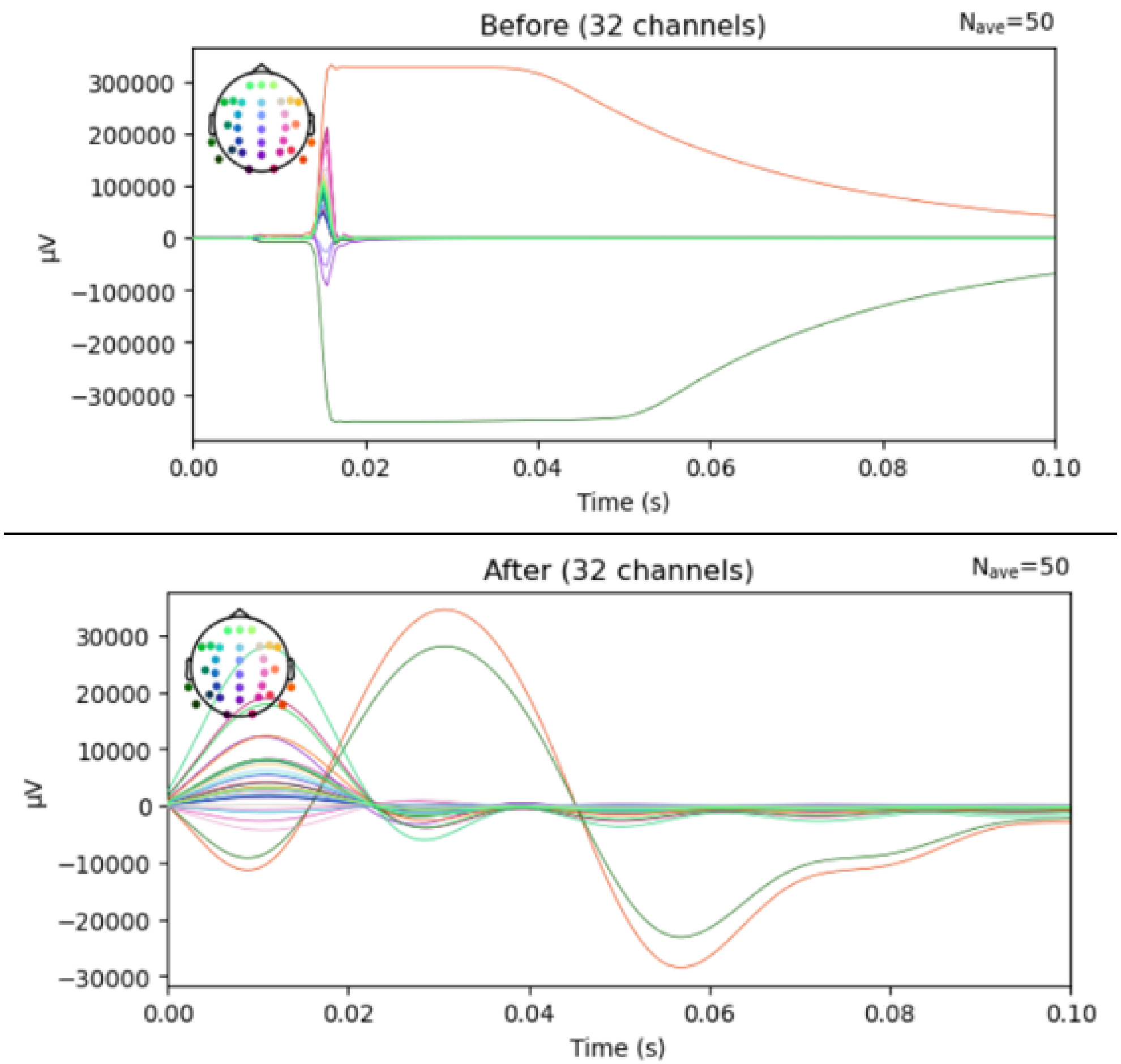

Figure 6 Pure SSP without SIR

Figure 6 shows the EEG signals before and after SSP correction. Before artifact correction, the EEG recordings were dominated by large-amplitude transient components, particularly around 0.01–0.02 s after the TMS pulse. After applying SSP, the large transient artifact was substantially reduced, and more oscillatory EEG patterns became visible across channels. However, several channels still exceeded the typical physiological EEG amplitude range of approximately 5–8 µV, suggesting the presence of residual artifacts and possible signal distortion. Therefore, while SSP effectively suppressed the dominant TMS-related artifact, it did not fully restore clean neural activity in the absence of SIR-based source reconstruction.

### 3.5 Source-estimate Utilizing Noise-Discarding (SOUND) algorithm

Figure 7 illustrates the outcome of the SOUND algorithm. After rejecting noise at the source level and projecting the cleaned estimates back to the sensors, all 28 channels are confined within a ±10 μV range. The early TMS pulse artifact remains visible, but the ringing patterns that followed it have disappeared. Across channels, the signals exhibit a similar trend, indicating that SOUND effectively removed inconsistent artifacts while preserving neural activity.

These results demonstrate that the algorithm can suppress residual neural signal noise, TMS-related artifacts, and other noise sources while retaining the main structure o the underlying cortical response. However, because a true physiological ground truth is unavailable, the SOUND-cleaned data should not be interpreted as an absolute representation of clean EEG activity. Instead, it can be considered a carefully preprocessed reference dataset for evaluating artifact removal performance and supporting future automated cleaning strategies.

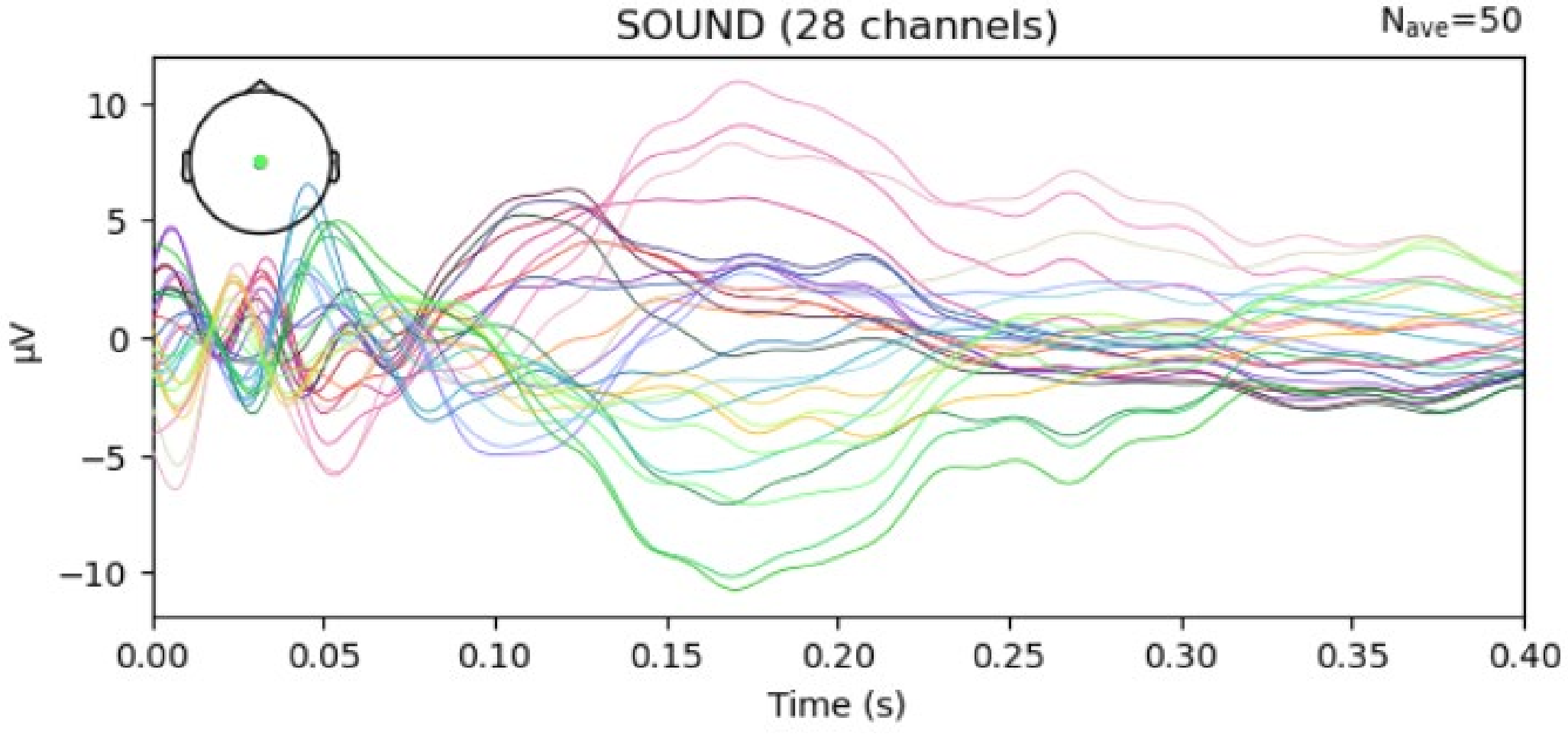

Figure 7 SOUND with 5 iterations with 0.3 lambda value

## 4 Results

### 4.1 Artifacts sources

1. TMS-related artifacts
    a. Pulse Artifact: A brief, high-amplitude electrical interference occurs after TMS delivery, and lasts for 1ms. It does not reflect actual neural activity
    b. Step Response: A transient electrical disturbance lasting approximately 5–7 milliseconds following TMS delivery. It does not reflect brain activity.
    c. Cranial Muscle Artifact: physiological artifact, EMG 5-10ms, 110-140hz

d. Decay Artifact: It occurs approximately 10–20 milliseconds post-TMS and can persist for up to 100 milliseconds.
e. Recharging Artifact: Arise due to the TMS stimulator's recharging state between pulses

2. Channel-wise noise

Channel-wise noise arises from variability in individual EEG electrodes, such as impedance fluctuations due to poor electrode-skin contact. This noise can lead to inconsistent signal quality across channels.

3. Ocular signals

Artifacts from ocular signals, such as eye blinks and eyeball movements. These ocular artifacts typically present as low-frequency, high-amplitude deflections.

4. Peripheral-evoked potentials

These environmental artifacts originate from external sources such as electrical noise generated by equipment, and constant-frequency interference emitted by LED lighting systems. Due to their unpredictability, additional measures for effective filtering are required.

### 4.2 Time frequency response

Time-frequency response (TFR) analysis is computed to study the dynamic activity of brain signals in response to TMS stimulation. This approach allows for the characterization of how neural oscillations evolve over time across different frequency bands, providing insight into both the timing and spectral content of cortical responses. By decomposing the EEG signal into time-resolved frequency components, TFR captures transient bursts of activity.

The first time-frequency representation (TFR), computed after the second Infomax ICA, still contains a lot of activity from artifacts. A strong vertical band of power appears at time zero, covering the entire 4–40 Hz range. This is caused by leftover muscle activity and coil ringing from the TMS pulse. The broadband noise from these artifacts hides weaker brain signals, and the expected beta rebound is barely visible. (Cannon et al., 2014) Because of these lingering artifacts, the TFR looks noisy and is hard to interpret. This shows the importance of applying an extra cleaning step like the SOUND algorithm to recover meaningful brain activity.

The second time–frequency representation (TFR), generated after the SOUND algorithm, is markedly cleaner. It reveals a pronounced post-stimulus beta rebound which is a transient increase in 15–30 Hz power that reflects the closure of the sensorimotor gating window once the immediate cortical response subsides. By suppressing broadband artifacts and enhancing the signal-to-noise ratio, SOUND allows genuine oscillatory activity to dominate over residual noise, yielding a more interpretable depiction of underlying neuronal rhythms.

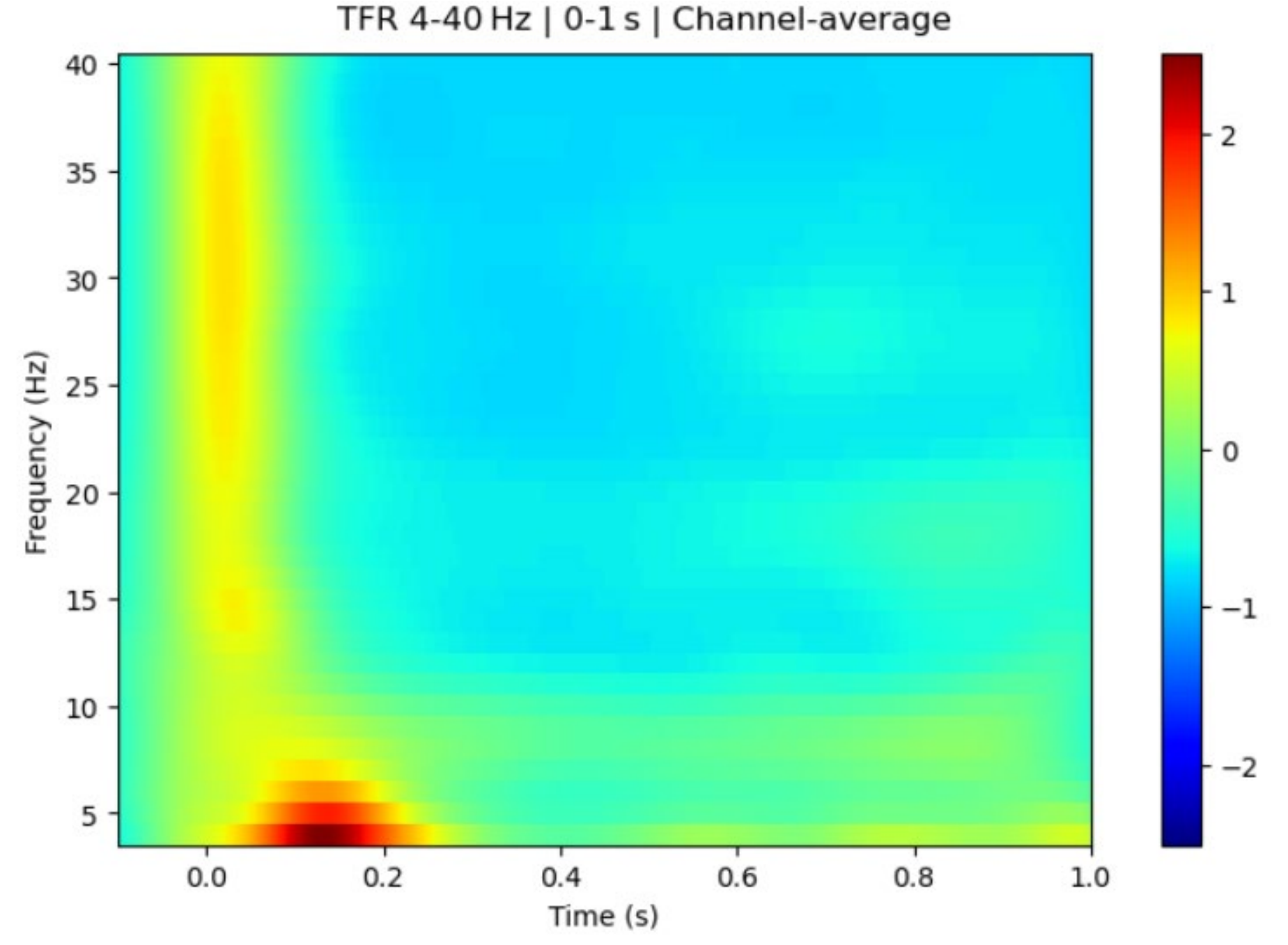

Figure 8 Second round of ICA

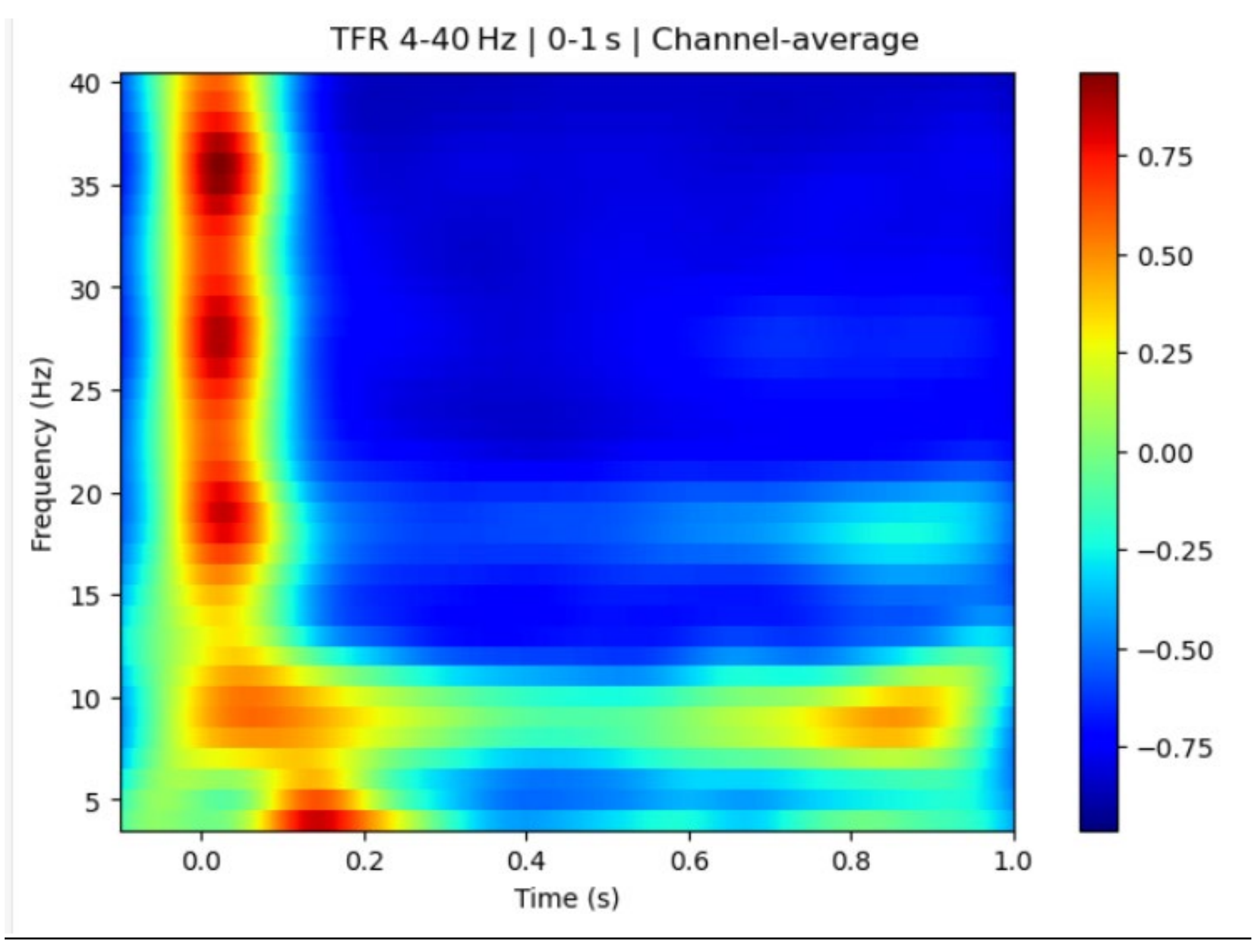

Figure 9 SOUND

**4.3 SOUND Computation**

To greatly improve the computation efficiency of the SOUND algorithm, Mutanen et al. recommend compressing lead field L with a singular-value decomposition (SVD) before the Wiener inversion. The matrix is compressed from 30 channels* 361984 to 32 channels *

15000 data point. After a signal trial cleaning of SOUND, there are several features we noticed

1) Amplitude compression

The traces now sit within roughly ±5 μV. That is an order of magnitude lower than the raw post-TMS signal, showing that SOUND has removed the large-amplitude coil artefact and much of the broadband muscle noise.

2) High overlap of channels

All colored traces are almost completely overlapping, inter-channel variance has been equalized, so what remains is the shared cortical activity with a small amount of uncorrelated sensor noise.

3) Preserved fast fluctuations

Temporal details have been retained.

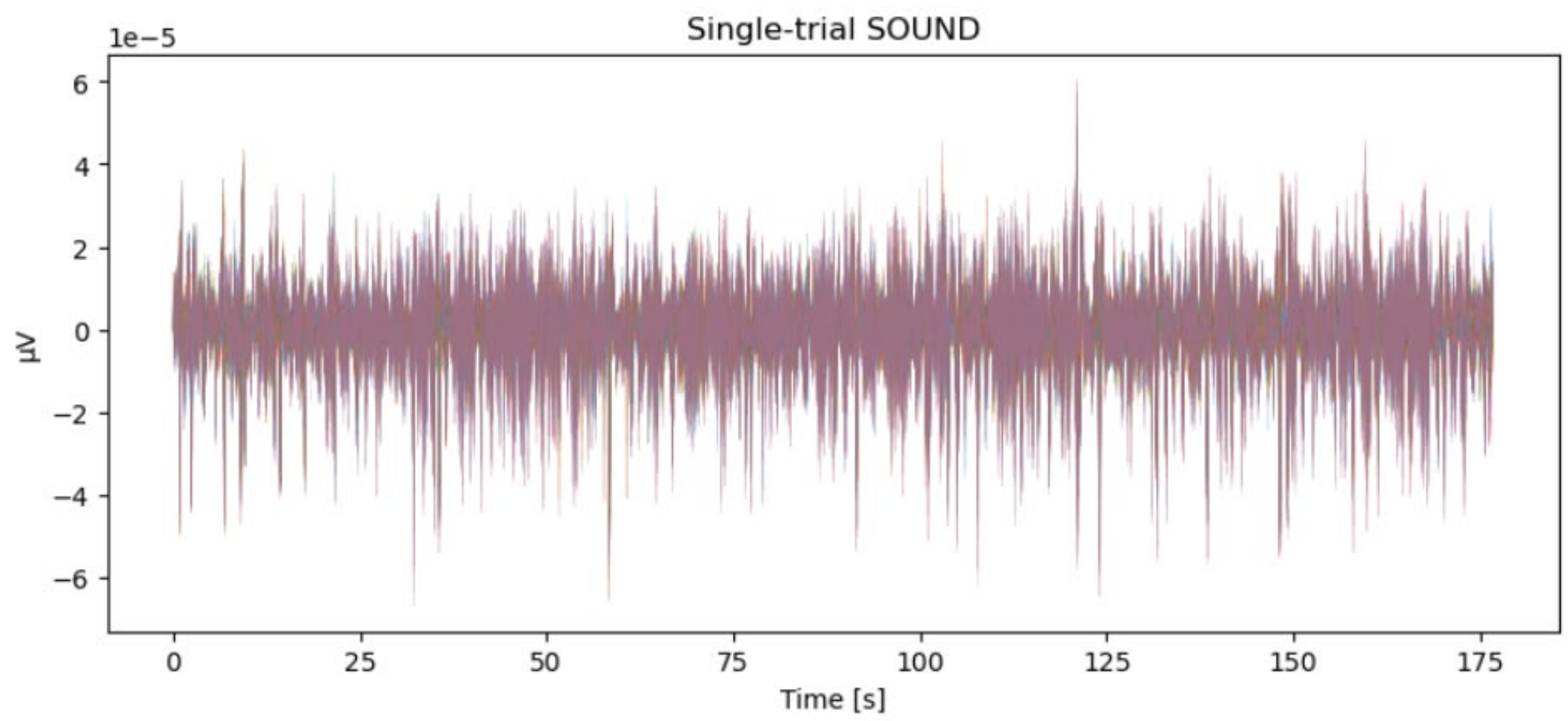

Figure 10 Single trial SOUND signals overlay

## 5 Conclusions and Future Work

This study achieved two milestones: 1) Reference dataset creation - a carefully preprocessed and artifact-cleaned TMS-EEG dataset was established to support the validation of future artifact removal algorithms. 2) Python SOUND implementation - Verified performance at the current stage comparable to MATLAB TESA version.

While ICA and SSP effectively reduce high-amplitude artifacts, they either distort spatial patterns (SSP) or risk removing brain signals with overlapping topographies (ICA). SOUND, by working in the source domain, better preserves neural activity while suppressing noise. We established a validated TMS–EEG cleaning pipeline and benchmark dataset. This research also evaluated two widely used TMS-EEG artefact-removal pipelines, i.e., SSP-SIR and SOUND. Since a true physiological ground truth is unavailable, the cleaned dataset can be interpreted as a carefully preprocessed reference dataset, providing a benchmark for future algorithm development and comparison of cleaning strategy. There are also FAST SOUND and real-time SOUND algorithms derived from the original SOUND. One possible direction is to develop a unified SOUND framework suitable for real-time artifact cleaning.

Integration with BCI research is an important long-term goal, as we aim to embed TMS-EEG within a larger brain-computer interface framework. Towards this objective, future studies will explore phase estimation models tailored to symptom provocation paradigms and investigations of synaptic plasticity. This direction may enhance the characterization of cortical dynamics and expand the clinical and research applications of TMS-EEG.